# Invertible Coarse Graining with Physics-Informed Generative Artificial Intelligence

Jun Zhang[1,†], Xiaohan Lin[2], Weinan E[3,4] and Yi Qin Gao[1,2]

*[1] Changping Laboratory, Beijing 102200, China*
*[2] Beijing National Laboratory for Molecular Sciences, College of Chemistry and Molecular Engineering, Peking University, Beijing 100871, China*
*[3] AI for Science Institute, Beijing, China*
*[4] Center for Machine Learning Research and School of Mathematical Sciences, Peking University, Beijing 100871, China*

Correspondence should be sent to: † jzhang@cpl.ac.cn (Jun Zhang)

## Abstract

Multiscale molecular modeling is widely applied in scientific research of molecular properties over large time and length scales. Two specific challenges are commonly present in multiscale modeling, provided that information between the coarse and fine representations of molecules needs to be properly exchanged: One is to construct coarse grained models by passing information from the fine to coarse levels; the other is to restore finer molecular details given coarse grained configurations. Although these two problems are commonly addressed independently, in this work, we present a theory connecting them, and develop a methodology called Cycle Coarse Graining (CCG) to solve both problems in a unified manner. In CCG, reconstruction can be achieved via a tractable deep generative model, allowing retrieval of fine details from coarse-grained simulations. The reconstruction in turn delivers better coarse-grained models which are informed of the fine-grained physics, and enables calculation of the free energies in a rare-event-free manner. CCG thus provides a systematic way for multiscale molecular modeling, where the finer details of coarse-grained simulations can be efficiently retrieved, and the coarse-grained models can be improved consistently.

# I. Introduction

Multiscale modeling is critical in various fields of scientific research including physics, chemistry, biology, materials and engineering. For many-particle Hamiltonian systems, coarse-graining of the microscopic model can drastically simplify the representation of the physical system. Specifically, in molecular simulations [1] interactions between particles are described by the potential energy $U(\mathbf{R})$ which is a function of the positions of particles (denoted by $\mathbf{R}$). Since $\mathbf{R}$ can be very high dimensional, the corresponding energy landscape $U(\mathbf{R})$ is usually rugged with many local traps. As a result, fine grained (FG) or first-principle based simulations, including all-atom and ab initio molecular dynamics (MD), are known to suffer from limited accessible time and length scales.

One solution against this issue is to extract slowly changing CG variables of the system $\mathbf{s} = s(\mathbf{R})$, and build CG models accordingly. For example, in a widely used linear coarse graining protocol, groups of atoms or particles from a fine-grained model are bundled into single beads [2-3], thus eliminating the microscopic degrees of freedom (DoFs) that are not essential to resolve structural features above a certain length scale. CG variables can also be non-linear functions of $\mathbf{R}$, which are more often termed as collective variables [4-6] and widely investigated in the context of enhanced sampling [7-9].

In molecular simulations, the CG potential $F(\mathbf{s})$ should satisfy the *thermodynamic consistency principle* (Eq. 1), which states that $F(\mathbf{s})$ should reproduce the marginalized Boltzmann distribution of the CG variables $\mathbf{s}$ given the FG potential $U(\mathbf{R})$,

$$p(\mathbf{s}) = \frac{\int e^{-\beta U(\mathbf{R})} \delta\big(\mathbf{s} - s(\mathbf{R})\big) d\mathbf{R}}{\int e^{-\beta U(\mathbf{R})} d\mathbf{R}} = \frac{e^{-\beta F(\mathbf{s})}}{Z} \tag{1a}$$

$$F(\mathbf{s}) = -\frac{1}{\beta}\left[\log p(\mathbf{s}) + \log Z\right] \tag{1b}$$

where $\beta$ is the inverse temperature, $\delta$ denotes the Dirac-$\delta$ function and $Z = \int e^{-\beta U(\mathbf{R})} d\mathbf{R}$ is the partition function. Under this setting, the CG potential $F(\mathbf{s})$ becomes a simplified description of the original thermodynamic system and is usually much smoother than the original energy landscape $U(\mathbf{R})$. Consequently, CG simulations performed under $F(\mathbf{s})$ are generally much faster, hence, can reach larger time and length scales that are inaccessible to the fine-grained models [3], despite that the finer details of the system are lost. Various approaches have been developed to parametrize CG models satisfying Eqs. (1-2), such as Iterative Boltzmann inversion [10], force matching [11], and relative entropy [12], etc. Some recent improvement along this line includes the deployment of artificial neural networks to augment the expressivity and complexity of the CG potentials [13-14], as well as the use of generative and reinforcement learning to boost the training efficiency of CG models in lack of FG simulation samples [15-16].

However, in many applications, one may need to simulate large systems with long correlation time in sufficiently fine details. Examples include the studies of the interaction between biological macromolecules [17], and the local structure of polymers at the interface with solid surfaces [18], etc. In these cases, a reconstruction approach, which will be referred to as *fine-grained reconstruction* (FGR) hereafter, is needed in order to retrieve or reconstruct reasonable FG structures consistent with the given CG model, and it has been noted that reconstruction is an integral part for systematical multiscale modeling [19]. Unfortunately, FGR has not been as well studied as CG, partly due to the ill-posed definition of FGR provided that mathematically $s(\mathbf{R})$ is usually a non-invertible function. In other words, for a given CG rule $s(\mathbf{R})$, there exists one unique way to map one FG structure $\mathbf{R}$ into one CG structure $\mathbf{s}$. In contrast, many different FG configurations can be assigned to the same CG structure, and it usually remains *ad hoc* to choose the “correct” one. Due to this difficulty, existing FGR methods, such as random mapping and position-restrained molecular dynamics (MD), commonly rely on system-specific knowledge and manually engineered fragment libraries [20-23], hence, are limited in scopes and lack of generalizability. Worse still, in most studies, CG and FGR problems were treated in a disconnected manner, and there lacks a unified theory to establish the relation between these two deeply coupled problems.

In this work, we formally formulate FGR as a probabilistic learning problem, and demonstrate that the FGR problem can be systematically solved by means of deep generative learning as an optimization problem. Moreover, we also draw a mathematical relation between the CG and FGR problems, and develop Cycle Coarse Graining (CCG), a general-purposed approach to performing multiscale modeling for molecular simulations: Based on machine learning, CCG delivers a tractable solution to the FGR problem. In turn, it also provides a rare-event-free way to perform coarse graining or calculate free energies, and allows efficient equilibrium sampling of rare events governed by the CG variables.

# II. Theory & Methods

## 1. Fine-grained reconstruction with thermodynamic consistency

In molecular simulations, given the coarse-graining function $s(\mathbf{R})$, any FG structure $\mathbf{R}$ that satisfies $s(\mathbf{R}) = \mathbf{s}$ can be considered as a valid reconstruction for the given CG structure $\mathbf{s}$. The distribution of all such valid reconstructions can be written in the form of a conditional probability (Eq. 2),

$$\mathbf{R} \sim p(\mathbf{R}; \mathbf{s}) \tag{2}$$

Equation (2) provides a general statistical viewpoint to the FGR problem, and exiting FGR methods can be categorized according to how they choose and sample from $p(\mathbf{R}; \mathbf{s})$. We then formally decompose the FGR problem into two sub-problems: i) defining a reasonable reconstruction distribution $p(\mathbf{R}; \mathbf{s})$, and ii) efficiently drawing samples from $p(\mathbf{R}; \mathbf{s})$. For the first task, a natural selection of $p(\mathbf{R}; \mathbf{s})$ is given by Eq. 3,

$$p(\mathbf{R}; \mathbf{s}) = \frac{e^{-\beta U(\mathbf{R})} \delta\big(\mathbf{s} - s(\mathbf{R})\big)}{Z(\mathbf{s})} \tag{3}$$

where $Z(\mathbf{s}) = \exp\big(-\beta F(\mathbf{s})\big)$ is the marginal partition function given a reference $\mathbf{s}$. In terms of physics, $p(\mathbf{R}; \mathbf{s})$ defined in Eq. (4) allows a CG structure to be reconstructed into thermodynamically favourable FG structures according to the Boltzmann distribution. We thus name Eq. 3 as the *thermodynamic consistency principle* for FGR, in analogy with the thermodynamic consistency principle for CG (Eq. 1).

The remaining task is to draw samples according to Eq. 3. Conventionally, this is done by restrained or targeted MD [6, 24]. However, after applying the restraints (typically in harmonic forms), computing the partition function in Eq. 3 becomes intractable. Alternatively, sampling according to Eq. 3 can be translated as a conditional generative learning problem, which has been intensively investigated by the machine learning community over topics like image super-resolution [25-26]. Therefore, we may turn the sampling problem into a more tractable optimization task and employ proper deep learning techniques to achieve this goal [27].

## 2. Deep generative learning for fine-grained reconstruction

In line with deep generative learning such as generative adversarial networks (GAN) [28], we start with a "generator", e.g., a neural network model parametrized by $\theta$ which generates samples $\mathbf{R}$ according to an optimizable generative distribution $q_\theta(\mathbf{R}; \mathbf{s})$. Usually sampling from $q_\theta(\mathbf{R}; \mathbf{s})$ is done via the re-parametrization trick [28-29],

$$\mathbf{R} = f_\theta(\mathbf{z}; \mathbf{s}) \Leftrightarrow \mathbf{R} \sim q_\theta(\mathbf{R}; \mathbf{s}) \tag{4a}$$

$$q_\theta(\mathbf{R}; \mathbf{s}) \triangleq \int \delta\big(f_\theta(\mathbf{z}; \mathbf{s}) - \mathbf{R}\big) p(\mathbf{z}) d\mathbf{z} \tag{4b}$$

where $f_\theta$ is a function transforming a random variable $\mathbf{z}$ into a configuration sample $\mathbf{R}$, and $\mathbf{z}$ usually comes from a tractable prior distribution (or base distribution) like standard normal.

We then aim to tune the model parameters $\theta$ so that the resulting $q_\theta$ is identical or close enough to the target reconstruction distribution $p(\mathbf{R}; \mathbf{s})$ defined in Eq. 3. To achieve this goal, we can minimize a strict divergence, for example, Kullback-Leibler divergence $D_{\mathrm{KL}}$ (Eq. (S1); see Supplemental Texts for more details), between the generative distribution and the reconstruction distribution. As in variational inference [30], $q_\theta$ can be optimized by minimizing $D_{\mathrm{KL}}\big(q_\theta(\mathbf{R};\mathbf{s})||p(\mathbf{R};\mathbf{s})\big)$ following the gradient in Eq. 5,

$$\nabla_\theta \mathcal{L}_{\mathrm{PI}}(\theta) = \nabla_\theta D_{\mathrm{KL}}(q_\theta||p) = \mathbb{E}_{\mathbf{R}\sim q_\theta}[\nabla_\theta \log q_\theta(\mathbf{R};\mathbf{s}) + \beta\nabla_\theta U(\mathbf{R};\mathbf{s})] \tag{5}$$

Note that computing this gradient can be done without calculating the partition function $Z(\mathbf{s})$, thus forgoing sampling from $U(\mathbf{R})$. Given access to a potential energy function $U(\mathbf{R})$ such as a force field, optimization of Eq. 5 can be done without data, so we call Eq. 5 the *physics-informed* training objective, $\mathcal{L}_{\mathrm{PI}}$, (see more details in SI).

On the other hand, if samples $\{\mathbf{R}_i\}_{i=1,\ldots N}$ are drawn via FG simulations, yielding a paired dataset $\mathcal{D} = \big\{\big(\mathbf{R}_i; s(\mathbf{R}_i)\big)\big\}$, we can also minimize the reversed KL divergence $D_{\mathrm{KL}}\big(p(\mathbf{R};\mathbf{s})||q_\theta(\mathbf{R};\mathbf{s})\big)$, which is called the *data-driven* training objective (Eq. 6), $\mathcal{L}_{\mathrm{DD}}$, and equivalent to maximum likelihood estimation (MLE), to optimize the generative distribution,

$$\nabla_\theta \mathcal{L}_{\mathrm{DD}}(\theta) = \nabla_\theta D_{\mathrm{KL}}(p||q_\theta) = \mathbb{E}_{\mathbf{R}\sim p}[-\nabla_\theta \log q_\theta(\mathbf{R};\mathbf{s})] \approx \mathbb{E}_{\mathbf{R}\in\mathcal{D}}[-\nabla_\theta \log q_\theta(\mathbf{R};\mathbf{s})] \tag{6}$$

Many options of deep generative neural networks, including flow-based and diffusion-based models [29, 31-32], can be used as $q_\theta$ and optimized with respect to the physics-informed or (and) data-driven objectives. More details about the suitable generative models can be found in Supplemental Texts.

The remaining issue is how to impose the constraint $s(\mathbf{R}) = \mathbf{s}$ when optimizing $q_\theta$ (or $f_\theta$). If $s(\mathbf{R})$ is a linear function, as in most particle-based CG models [2, 33], the constraint can be rigorously preserved by means of substitution of variables. While in a more general case where $s(\mathbf{R})$ is non-linear, we can approximate $\delta\big(\mathbf{s} - s(\mathbf{R})\big)$ in Eq. 3 using smooth functions, although Lagrangian methods [34-35] can also be used. Taking physics-informed training as example, when a Gaussian kernel is used to approximate the Dirac $\delta$ function, the gradient of the loss function becomes,

$$\nabla_\theta \mathcal{L}_{\mathrm{PI}} = \mathbb{E}_{\mathbf{R}\sim q_\theta}[\nabla_\theta \log q_\theta(\mathbf{R};\mathbf{s}) + \beta\nabla_\theta U(\mathbf{R};\mathbf{s}) + \lambda\|\mathbf{s} - s(\mathbf{R})\|^2] \tag{7}$$

where $\lambda$ characterizes the bandwidth of the Gaussian kernel. In this way the constraint is relaxed into a restraint, and the hyper-parameter $\lambda$ can be regarded as a regularization factor which is reminiscent of the force constant defining a harmonic restraint potential applied in the restrained MD simulations. We call this restraint term *consistency regularization* or *cycle loss*, and remark that it has other interpretations in related machine learning literature. For example, a similar term is incorporated by Cycle GAN for unsupervised image-to-image translation [36], and used for mutual information maximization which helps overcome the mode-dropping issue in the studies of GANs [37].

## 3. Cycle coarse graining

An important relation can be drawn between CG and FGR on the basis of the two thermodynamic consistency principles, namely, Eq. 1 and Eq. 3. Given a deterministic CG rule $s(\mathbf{R})$, the Boltzmann distribution of the FG potential $U(\mathbf{R})$, $p(\mathbf{R}) \propto e^{-\beta U(\mathbf{R})}$, can be factorized in the form of Eq. 8,

$$p(\mathbf{R})d\mathbf{R} = p\big(\mathbf{R}, s(\mathbf{R})\big)d\mathbf{R} = \int p(\mathbf{s})p(\mathbf{R};\mathbf{s})d\mathbf{s} \tag{8}$$

where $p\big(\mathbf{R}, s(\mathbf{R})\big)$ denotes the joint distribution of $\mathbf{R}$ and $s(\mathbf{R})$, $p(\mathbf{s})$ is given by Eq. 1 and $p(\mathbf{R};\mathbf{s})$ by Eq. 3, respectively. Equation 8 shows that, to investigate the equilibrium properties of the FG model, instead of directly sampling from $p(\mathbf{R})$ which is typically time-consuming, one can perform *factorized sampling*: First sample at a coarse-grained scale according to $p(\mathbf{s})$, then perform FGR according to $p(\mathbf{R};\mathbf{s})$ and retrieve finer details. Factorized sampling thus corresponds to a general multi-scale sampling approach. Intuitively, Eq. 8 would not hold unless that the motions of CG variable $s(\mathbf{R})$ can be decoupled with other DoFs, echoing the common practice that only the most slowly varying DoFs shall be chosen as CG variables. One intriguing advantage of factorized sampling lies in its ability to "interpolate" or "extrapolate" along the CG variables. Consider that $s(\mathbf{R})$ is a characteristic chemical reaction coordinate, Eq. 8 then allows us to explore the reaction coordinate in a rare-event-free manner, and can be further boosted by existing enhanced sampling methods [38-41]. Although such interpolated trajectories can do not necessarily correspond to the true reaction pathways, however, many approaches can be implemented to retrieve the correct kinetics or transition pathways given the reconstructed free energy profile [42-43].

Finally, we notice that solving the FGR problem defined in Eq. 3 simultaneously yields a novel solution to the CG problem in Eq. 1. Specifically, given a bijective generator $f_\theta(\mathbf{z};\mathbf{s})$ corresponding to a generative distribution $q_\theta(\mathbf{R};\mathbf{s})$ (Eq. 4) which minimizes $D_{\mathrm{KL}}(q_\theta||p)$ or $D_{\mathrm{KL}}(p||q_\theta)$, the *variational free energy* $F_\theta(\mathbf{s})$ in the form of Eq. 9 is a good approximation to the ground-truth free energy $F(\mathbf{s})$ in Eq. 1b,

$$\beta F_\theta(\mathbf{s}) = \mathbb{E}_{\mathbf{z}\sim p(\mathbf{z})}\big[\log f_\theta(\mathbf{z};\mathbf{s}) + \beta\nabla_\theta U\big(f_\theta(\mathbf{z};\mathbf{s})\big)\big] \tag{9a}$$

$$p_\theta(\mathbf{s}) \triangleq \exp\big(-\beta F_\theta(\mathbf{s})\big) \tag{9b}$$

where $F_\theta(\mathbf{s})$ is an upper bound to $F(\mathbf{s})$ (see Supplemental Texts for the derivation of Eq. 9). As a result, we can approximate the free energy or construct CG potential function of $\mathbf{s}$ through an optimized FGR generator $f_\theta$ according to Eq. 9. The assembled training and inference protocol of CCG is summarized in Supplemental Information and Algorithm S1.

In summary, by solving the FGR problem, we can obtain a generative model $q_\theta(\mathbf{R};\mathbf{s})$ which reconstructs FG structures according to the CG variables (Eq. 4). On the other hand, the optimized FGR model $q_\theta$ in turn gives rise to a CG potential $F_\theta(\mathbf{s})$ (Eq. 9), without calculating the mean forces or sampling the rare events explicitly. Finally, combining $F_\theta$ and $q_\theta$, we can perform efficient multiscale sampling over a complex $U(\mathbf{R})$ in a factorized fashion (Eq. 8). This workflow forms a cycle between CG and FGR, hence, is named altogether as Cycle Coarse Graining (CCG), and provides a novel and self-consistent framework for multiscale molecular modelling.

# III. Experiments & Results

## 1. Benchmark cycle coarse graining on numerical potentials

We first benchmarked CCG on a 2-dimensional numerical model and illustrated how it works with different settings. The potential energy surface $U(x, y)$ of Tiwary-Berne model [44] is shown in Fig. 1a, which consists of three local minima. The optimal linear reaction coordinate $s = x\cos\alpha + y\sin\alpha$ (where $\alpha = 81.5^{\circ}$ is the skewing angle from $x$-axis) for this model was derived by Tiwary et al. [44] (dashed line in Fig. 1a), and computing potential of mean force (PMF, or free energy) along $s$ entails time-consuming simulations via either Monte Carlo or Langevin dynamics. Although being simple, this model is characteristic of many bio-physical systems composed of multiple metastable states and the interstate transitions are rare events. We thus treated $s$ as the CG variable and tried to reconstruct $(x, y)$ given the 1-dimensional $s$.

We first performed physics-informed training for FGR according to Eq. 5. We adopted FFJORD [45], one type of deep bijectors to model the generative distribution $q_\theta(x, y; s)$, and randomly sampled $s$ according to a uniform distribution. Since $s$ is linear function of $(x, y)$, we resorted to substitution-of-variables instead of consistency regularization, rigorously satisfying the constraint (see more details about training and model setup in Supplemental Information). Given the optimized $q_\theta$, we computed the variational free energy along $s$ according to Eq. 9. Figure 1b shows the resulting variational free energy profile learned by CCG, which has neither performed rare event sampling nor required transitions between metastable states. However, the accuracy is remarkably good compared to the reference PMF which is computed by integrating over the true potential energy surface, showing that CCG enables free energy calculation or coarse graining in a rare-event-free approach.

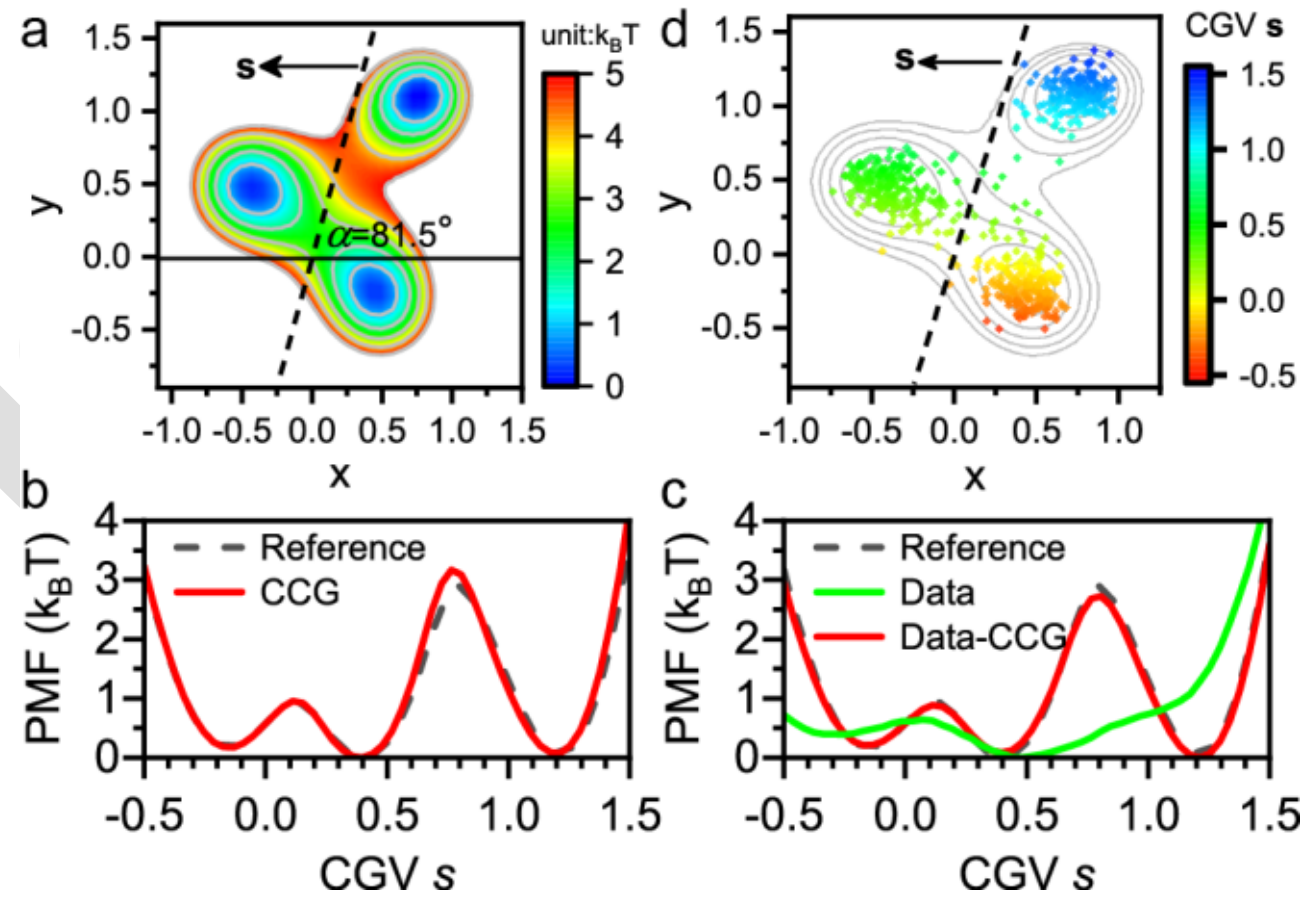

**Figure 1.** Illustration of cycle coarse graining (CCG) on a numerical model. **(a)** The potential energy surface (PES) of the model, shown in colored contour plot. The dashed line indicates the linear reaction coordinate **s** with a skewing angle $\alpha$. **(b)** The potential of mean force (PMF) computed by physics-informed CCG (red line) compared to ground-truth reference (black dashed line). **(c)** The potential of mean force computed by data-driven CCG (red line) compared to ground-truth reference (black dashed line). Distribution of the off-equilibrium data is also shown (green line). **(d)** Generated data through factorized sampling using physics-informed CCG model, colored according to the CG variable **s** during FGR. The PES is shown as grey contours in background.

We also benchmarked data-driven FGR (Eq. 6). To simulate better the real-world cases, we assumed that the accessible data of $(x, y)$ are distributed off equilibrium. To do so, we ran Langevin dynamics simulation with much higher temperature and collected the samples as shown in Fig. S1b, and the corresponding data distribution along $s$ is shown by green line in Fig. 1c which deviates dramatically from the reference as expected. Besides, we tested the strength of cycle loss ($\lambda$ in Eq. 7) ranging from $10^{-3}$ to 10, and found that training was robust with respect to the choice of $\lambda$. Particularly, even if the regularization strength was set to be rather weak ($\lambda = 10^{-3}$), the consistency regularizer still converged much faster than the generative objective, and achieved negligible errors after training (Fig. S1a), indicating that a small $\lambda$ would suffice for the relaxed FGR objective.

Given that the training data is off-equilibrium, we optimized FFJORD according to Eq. (S16), a modified version of Eq. 6 with reweighting trick, and applied correction to the variational free energy according to Zwanzig's free-energy perturbation theory [46] (Eq. (S17); see SI for more details). Figure 1c shows that the resulting free energy profile along $s$ agrees well with the reference, demonstrating that CCG can rescue free energy calculations even if the training data is distributed far off equilibrium.

To generate samples from $U(x, y)$, we can perform factorized sampling based on the CG potential $F_\theta(s)$ according to Eq. 8: We first conducted Monte Carlo sampling of $s$ according to $F_\theta(s)$, then performed FGR with an optimized $q_\theta(x, y; s)$. The generated samples from an energy-trained $q_\theta$ is shown in Fig. 1d, which conform to the correct Boltzmann distribution. Similar results were also obtained for models trained by data-driven objective (Fig. S1c).

Finally, we can generate fake trajectories connecting metastable states by interpolating or extrapolating the CG variable $s$, and we term this technique as "*trajectory interpolation*". Trajectory interpolation can be done simply by fixing the random variable $\mathbf{z}$ of the generative model $f_\theta(\mathbf{z}; \mathbf{s})$ (Eq. 4a) while varying the conditional variable $\mathbf{s}$ as desired, then recording the corresponding output of the model $\mathbf{R} = f_\theta(\mathbf{z}; \mathbf{s})$. By selecting different random variables $\mathbf{z}$ and performing trajectory interpolation for each $\mathbf{z}$, one can obtain an ensemble of fake trajectories. We showcased such a fake trajectory generated by interpolation in Fig. S1d, which connects all the three metastable states and passes through barriers between them, implying the possible application of this technique for investigating the transition states of chemical reactions.

## 2. Cascaded fine-grained reconstruction for proteins

We then applied CCG for a mini-protein chignolin [47]. To verify the generic applicability of CCG, we performed FGR for chignolin at two different scales. At the first scale, the root-mean-squared deviation (RMSD) of C$\alpha$ atoms from the native structure is treated as the 1-dimensional CG variable $\mathbf{s}$, and all the C$\alpha$ positions (denoted by $\mathbf{r}$) are reconstructed, giving rise to C$\alpha$ structures. Note that RMSD is a non-linear CG mapping, so reconstruction of C$\alpha$ structures from RMSD values is not straightforward using existing methods. One commonly adopted approach is targeted MD which samples possible protein structures under restraints of RMSD. However, targeted MD is very time-consuming, and depends severely on initial conditions and other hyperparameters like restraint strength; worse still, it can neither guarantee the diversity of reconstructed structures, nor the consistency with respect to the RMSD values. In contrast, we will show that CCG can efficiently reconstruct diverse FG samples of high quality and consistency.

We performed data-driven training of a generative $\mathbf{r} = f_{\theta_1}(\mathbf{z}; \mathbf{s})$ (Eq. 6) using all-atom trajectories contributed by Lindorff-Larsen et al. [48] and applied consistency regularization (see more training details in Supplemental Information). After training is done, we can generate C$\alpha$ protein structures at a certain RMSD by feeding a random variable $\mathbf{z}$ (drawn from standard normal distribution) and the specified $\mathbf{s}$ = RMSD to $f_{\theta_1}(\mathbf{z}; \mathbf{s})$. Figure 2 Row 1 shows the equilibrium free energy profile along **s** computed from the long MD simulations (Column a). Various reconstructed structures can be reconstructed at any specific **s** by means of CCG as shown in Row 2 in Figure 2. Particularly, we selected five milestones along the RMSD profile (Column **b** to **f** in Fig. 2 Row 1) at which the reconstructed structures were inspected. Three of the five states (Column **b** to **e**) correspond to unfolded states which exhibit large RMSD with respect to the native structure, one state (Column **f**) corresponds to folded structure, and one (Column **e**) resides nearby the "transition state" of folding. Figure 2 Row 2 shows a fake trajectory interpolated along the five selected milestones: When RMSD is relatively large, extended conformations are generated; the hairpin-like structure forms near the transition state and a folded structure is yielded when RMSD approaches zero. We also verified that the generated structures given by CCG can be aligned reasonably with a counterpart found in real MD simulations, further validating the quality of these reconstructed structures.

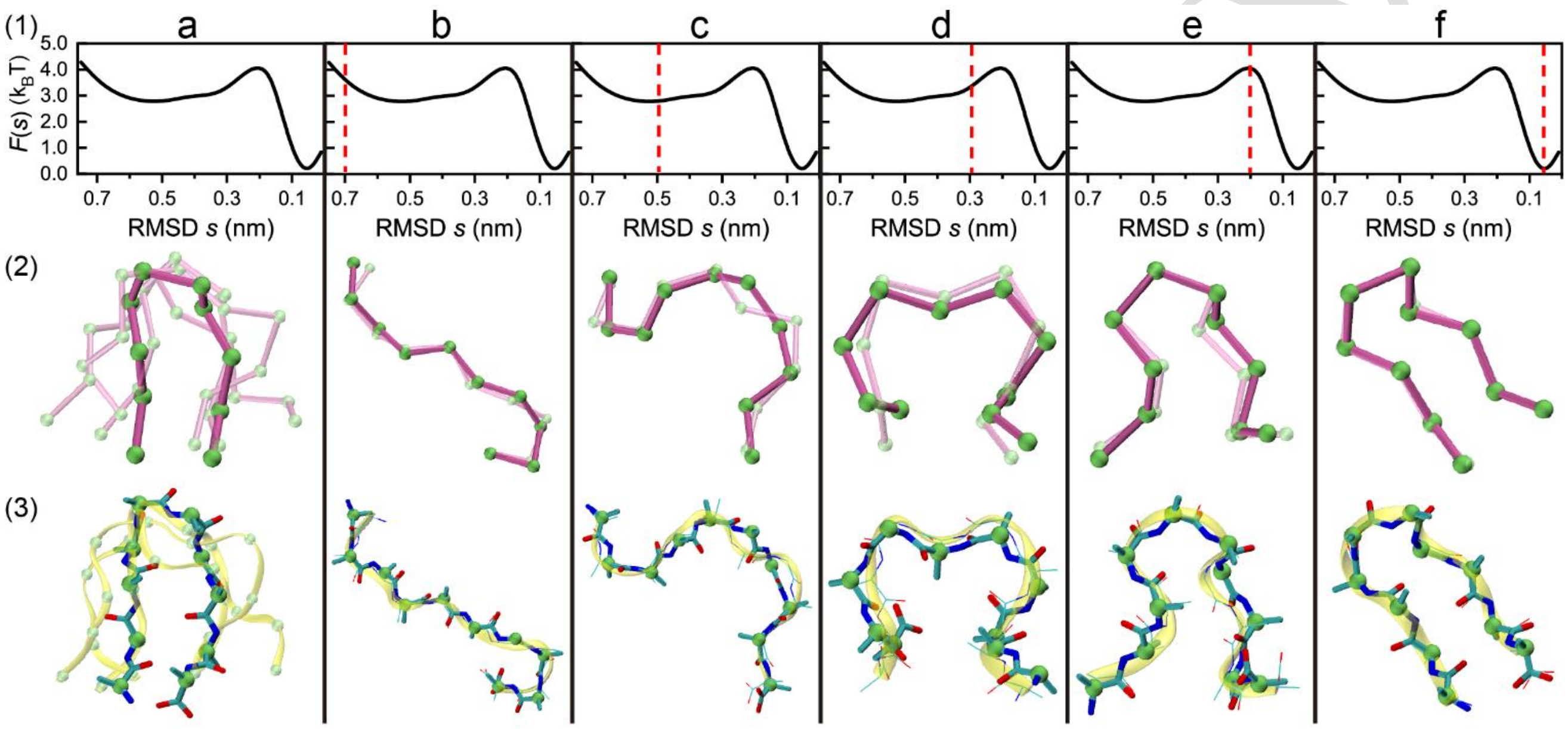

**Figure 2.** Multiscale fine-grained reconstruction for Chignolin. **Row 1: (a)** The PMF along the CG variable $s$, i.e., RMSD with respect to the native structure. **(b) to (f)** Five selected RMSD values at which the FGR results are inspected; **(b), (c)** and **(d)** correspond to unfolded states, **(e)** to the folding transition state and **(f)** is the folded state. **Row 2: (a)** C$\alpha$-structures are reconstructed given specified RMSD values. **(b) to (f)** show reconstructed C$\alpha$ structures (opaque magenta) by trajectory interpolation at corresponding RMSD in Row 1; The best aligned MD structure is shown in transparency. **Row 3: (a)** Structures with backbone heavy atoms (BH-structures) are reconstructed from C$\alpha$-structures. **(b) to (f)** show BH-structures reconstructed from the corresponding C$\alpha$-structures in Row 2; The best aligned MD structure is shown as transparent yellow ribbons.

Next, we performed CCG at another scale, where backbone structures (denoted by **R**) are reconstructed from a given C$\alpha$ structure **r**, with all backbone heavy atoms being added. Since this is a linear FGR problem, we trained the generative model $\mathbf{R} = f_{\theta_2}(\mathbf{z};\mathbf{r})$ according to the data-driven objective without applying consistency regularization. Now these two FGR models can be cascaded and generate FG structures at multiple scales given a certain RMSD: We first specified a RMSD value and generated C$\alpha$ structures **r** with $f_{\theta_1}(\mathbf{z};\mathbf{s})$, then generated backbone structures **R** with $f_{\theta_2}(\mathbf{z};\mathbf{r})$. Put together, backbone structures can be generated according to a certain RMSD. Following this way, backbone structures are generated corresponding to the selected **s** and **r**, which are shown in Fig. 2 Row 3. In Figure S2, we presented more reconstructed structures generated through cascaded trajectory interpolation. It can be found that when RMSD approaches zero, the generated structures aligned better with each other, showing convergence of conformations in the folded state. Conversely, when RMSD is large, the entropic effect dominates, and an increment of conformational diversity and flexibility is observed.

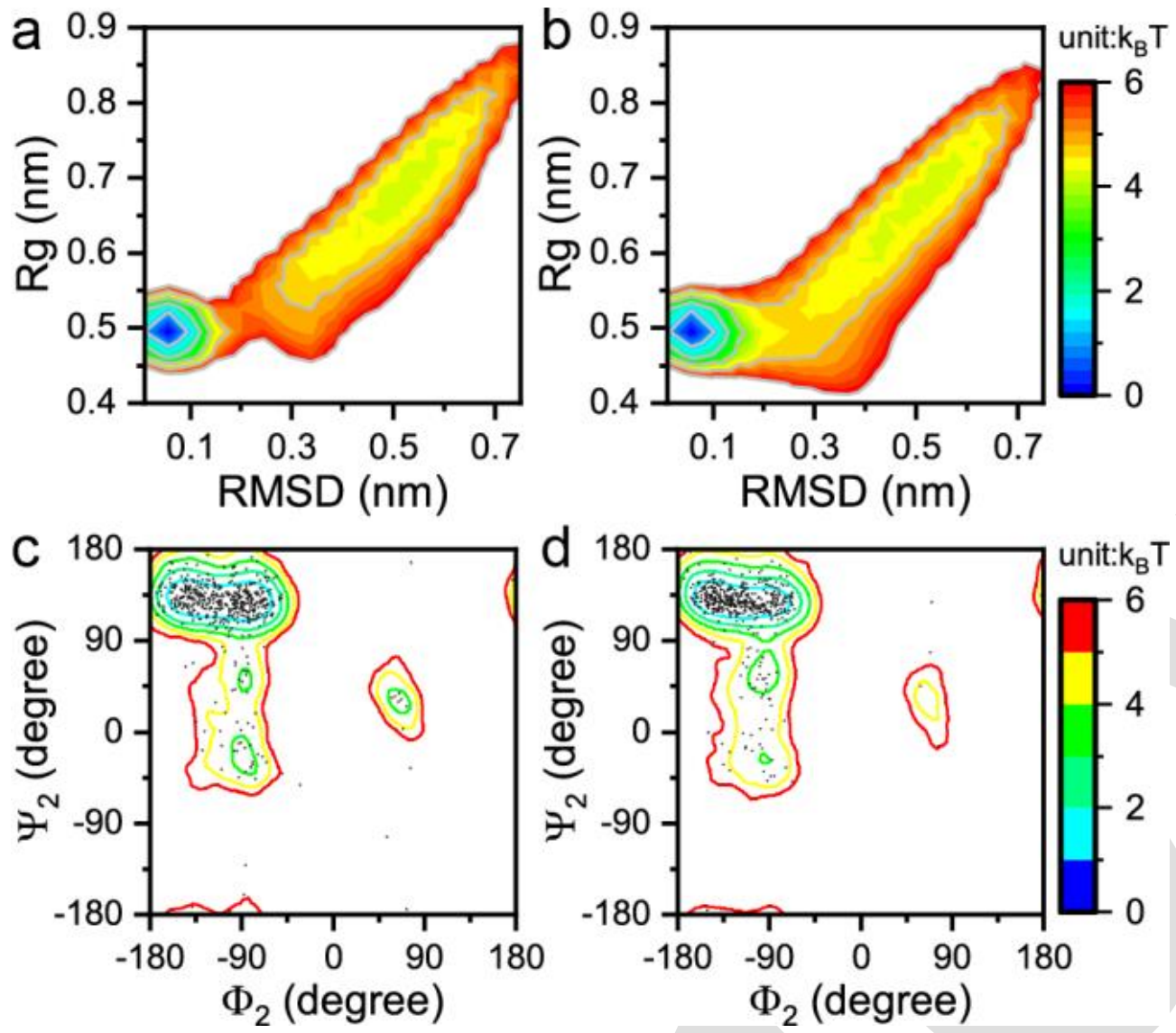

**Figure 3.** Conformational distribution of Chignolin through factorized sampling. The 2D free-energy plot for RMSD (with respect to native structure) and radius of gyration is drawn for MD samples **(a)** and factorized sampling **(b)**, respectively. The Ramachandra plot for the 2nd residue is drawn for MD samples **(c)** and factorized sampling (**d**), respectively.

It is also appealing to examine whether multiscale factorized sampling of protein backbone structures satisfy the thermodynamic consistency principle in Eq. 3. Because conformations of proteins are of particular research interest, we compared the distribution of RMSD and radius of gyration (both are global characterizations of protein conformations) of generated structures against samples from equilibrium all-atom MD simulations, as shown in Fig. 3a and 3b. Furthermore, we are often concerned with the detailed local structures of a protein, so we showed the Ramachandra plots of all the torsional angles for backbone structures generated by factorized sampling and compared them with samples from long MD simulations (Fig. 3d and 3e, Fig. S3). As Figure 3 shows, the conformations generated via factorized sampling reproduce those sampled via long MD simulations, proving the thermodynamic consistency of CCG. This implies that by virtue of CCG, knowing the PMF along a low-dimensional CG variable or collective variable like RMSD may suffice for obtaining more detailed conformational statistics of a protein.

## 3. Rare-event-free sampling of chemical reactions

Since trajectory interpolation circumvents sampling of rare events, it can be particularly useful in the studies of many biophysical processes involving rare transition events. As an important case, simulation of chemical reactions is often hindered by two major challenges: 1) chemical transitions are rare events due to high reaction barriers, thus a redundant number of single-point energy calculations are needed; 2) evaluation of the single-point energy involves quantum calculations which are slow and expensive. Recent advent of machine-learned potential [49-50] could help soothe the latter issue, by providing a surrogate potential to the quantum oracle which can be evaluated much faster. However, optimization of the surrogate potential requires training labels covering all the relevant reaction phase space with reference to the oracle [51], which entails exploration over the reaction coordinates and sampling of rare events. We will show that CCG can be employed to solve these issues effectively following an active learning workflow (Fig. 4) and the training protocol is summarized in Algorithm S2.

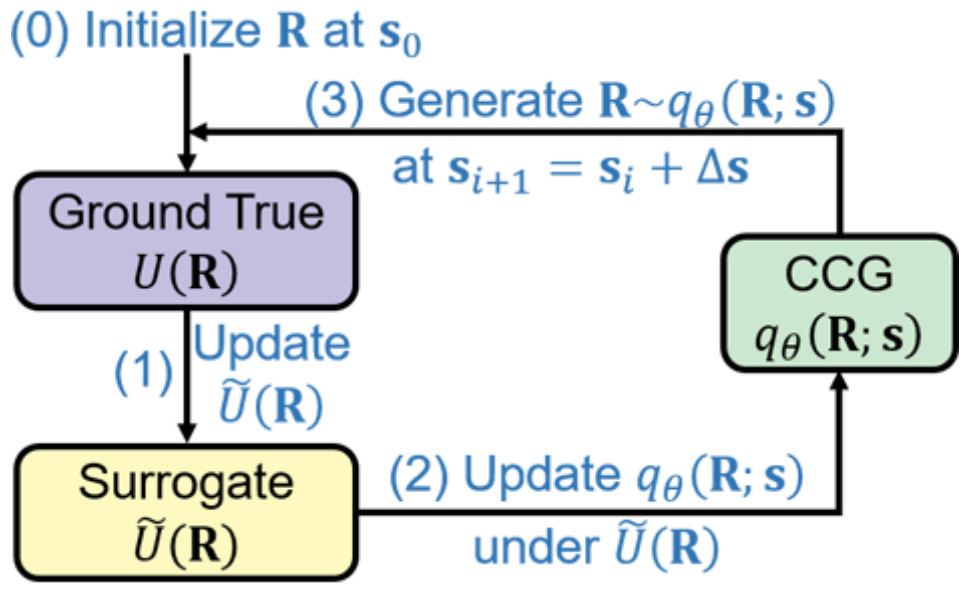

**Figure 4.** Rare-event-free workflow for active learning of chemical reactions via CCG. In one iteration, the all-atom structures $\mathbf{R}$, generated by either initialization or CCG, are fed to oracle quantum potential function and yield training labels $U(\mathbf{R})$. The surrogate machine-learned potential $\widetilde{U}(\mathbf{R})$ is updated by supervision of these labels. The CCG model $q_\theta(\mathbf{R}; \mathbf{s})$ is then updated based on $\widetilde{U}(\mathbf{R})$, and generates new samples by extrapolating the reaction coordinates $\mathbf{s}$, which are labeled during next iteration.

As Fig. 4 illustrates, given initial configurations $\mathbf{R}$, we first label them by calling the oracle $U(\mathbf{R})$, then train the surrogate function $\widetilde{U}(\mathbf{R})$ using these labels and update the CCG model $q_\theta(\mathbf{R}; \mathbf{s})$ with respect to $\widetilde{U}(\mathbf{R})$. This allows us to quickly obtain diverse samples along $\mathbf{s}$ without performing rare-event sampling. These samples are further queried by the oracle and serve as new labels for the surrogate potential. Hence, the CCG model $q_\theta$ and the surrogate potential $\widetilde{U}$ can be optimized in an alternating manner. When both models converge, the free energy profile along $\mathbf{s}$ as well as molecular configurations can be directly obtained from $q_\theta$ without performing MD simulations.

To examine whether CCG can help study chemical reactions in a computationally cheap and rare-event-free manner, we applied this strategy (Fig. 4) to investigate a textbook reaction, the substitution between $Cl^-$ and $CH_3Cl$ (Fig. 5a) which is known to undergo a typical $S_N2$ mechanism. The positions of three reactive atoms (Cl-, Cl and C) are selected as the CG variable $\mathbf{s}$, and we transformed the Cartesian coordinates into three internal coordinates (Fig. 5a), namely, the lengths of two C-Cl bonds ($d_1$, $d_2$) and the C-Cl-C reaction angle ($\alpha$), which suffice to describe the relative positions of these three atoms. We trained the surrogate potential as a function of all atom positions to approximate the reference atomic forces and energies, and employed a deep bijective model as $q_\theta(\mathbf{R}; \mathbf{s})$ which output all atom positions $\mathbf{R}$ including hydrogens (see Supplemental Information for more details of models).

After training, we can draw samples at given $\mathbf{s}$ and compute the variational free energy according to Eq. 9. The free energy surface (FES) is plotted at different reaction angles ($\alpha$) in Fig. 5b. Because this reaction is symmetric with respect to two Cl atoms, the FES appears symmetric as expected. At $\alpha = 180^\circ$ where C-Cl-C reside on a straight line, there is a unique saddle point across the FES residing at about $d_1 = d_2 \approx 2.45$ Å, indicating that bond breaking is synchronized with bond forming, consistent with a $S_N2$ mechanism. However, when we changed $\alpha$ to $135^\circ$, mimicking the reaction triggered by non-straight collisions (Fig. 5b), although the FES at product or reactant region does not change significantly, the saddle point region, which is key to the reactivity, is dramatically leveled up, excluding the possibility of such reaction pathways.

Next, we interpolated linearly between $d_1$ and $d_2$ with $\alpha$ fixed at $180^\circ$, and Fig. 5c shows the all-atom structures, which are generated by the optimized $q_\theta$, of the reactant, product, and transition state along an interpolated trajectory. Noteworthy, these structures exhibit authentic chemical details. For instance, the $CH_3Cl$ molecules in reactant or product both show a tetrahedron shape in agreement with a sp3 central carbon. However, for the transition state (TS) complex, the hydrogens become planar with respect to the central carbon, and they form trigonal bipyramid together with the two Cl atoms. To verify whether the chemical details are generally preserved as reaction proceeds, we performed factorized sampling and generated all-atom structures at different $\mathbf{s}$. The distribution of the angle $\alpha$ was computed based on these samples (Fig. 5d). It can be seen that, when $Cl^-$ is far away from $CH_3Cl$ (i.e., $d_2 > 4$Å), its orientation with respect to the other C-Cl bond is relatively arbitrary and isotropic. When $Cl^-$ approaches the reactive center (characterized by a shortened $d_2$ value), its orientation with respect to C-Cl bond becomes more restrained. When the TS complex is formed, the C-Cl-C angle becomes highly restrained and only $\alpha \approx 180^\circ$ is allowed (i.e., three atoms in a line). Similarly, we also showed the distribution of H-C-Cl angle ($\gamma$) in Fig. 5e. In reactant or product, angle $\gamma$ stays around $110^\circ$, corresponding to a sp3-hybridized carbon. But in the TS complex (defined as $d_1, d_2 \leq 2.55$ Å), this angle shifts to about $90^\circ$, indicating that Cl-C bond is orthogonal to the $CH_3$ planar, which also agree well with the chemical fact.

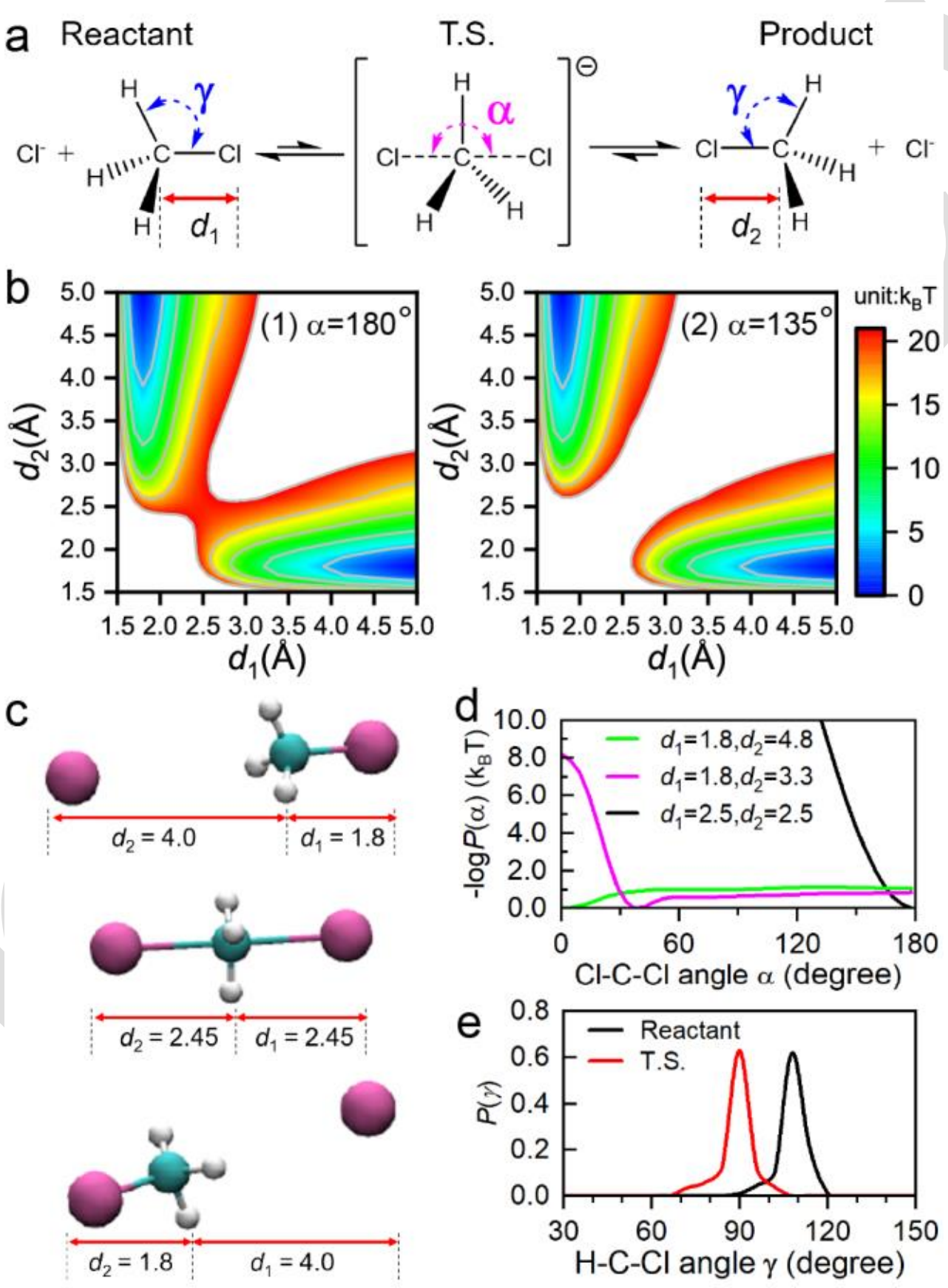

**Figure 5.** Rare-event-free sampling of $S_N2$ reaction: the substitution between $Cl^-$ and $CH_3Cl$. **(a)** Scheme of the reaction, where reactant, product and transition state complex (T.S.) is shown. Important structural descriptors are also illustrated, including the lengths of two C-Cl bonds ($d_1$ and $d_2$, respectively), Cl-C-Cl angle ($\alpha$) and H-C-Cl angle ($\gamma$). **(b)** Free-energy surface spanned by the two C-Cl bond lengths at different Cl-C-Cl reaction angles: $\alpha = 180^\circ$ (left), $\alpha = 135^\circ$ (right). **(c)** Snapshots of atomic structures through trajectory interpolation, including reactant (top), transition state (middle) and product (bottom); Carbons are colored by cyan, chlorides by magenta, and hydrogens by white. **(d)** Distribution of Cl-C-Cl reaction angle $\alpha$ at different reaction stages. **(e)** Distribution of H-C-Cl angle $\gamma$ in reactant (black) and transition state (red).

# IV. Concluding Remarks

Multiscale molecular modeling is very useful in molecular science where molecular properties at large time and lengths scales are of interest. But its application is limited by two long-standing challenges: One is to construct coarse-grained models by proper abstraction of fine-scaled models; the other is to restore finer molecular details given coarse-grained configurations. Although these two problems are commonly addressed independently, in this work, we presented a theory connecting them, and developed CCG to solve both problems in a consistent manner.

In CCG, we formulated fine-grained reconstruction as a probabilistic learning problem, and delivered a tractable solution to this task by means of machine learning. Through experiments we demonstrated that CCG is a reliable strategy for high throughput and high accuracy conversion of CG structures into FG ones for complex biophysical processes like protein folding and chemical reactions. Moreover, CCG provides a rare-event-free approach to coarse graining or free energy calculations. Specifically, trajectory interpolation enables fast exploration of the CG space that governs the slow motions of the system, whereas conventional MD simulations would be inefficient in sampling these transitions involving rare events. On the other hand, by separating the slow coarse-grained or collected motions and fast fine-grained degrees of freedom, factorized sampling leads to a rigorous multiscale approach to expediting the investigation of equilibrium properties of complex systems.

The methodology presented in this paper can be easily extended to other biophysical systems at different coarse-graining levels, such as biomacromolecules and polymers, where tuning atomistic details via human intervention is particularly labor-intensive. Besides, trajectory interpolation as well as factorized sampling makes it easy to marry other pathway-searching techniques like milestoning [52] and string method [42]. Although the generative models employed in this paper are all bijective, various quasi-invertible models [53-54] have been developed by the machine learning community recently, and they can help extending the application scope of CCG. In addition to thermodynamic consistency discussed throughout this paper, CG or FGR pursuing correct dynamics is also an active research field. Some recent efforts proposed machine learning approaches to extract dynamic information from molecular simulation trajectories [55-56], where CCG may also deliver helpful solutions, and we leave this direction to further studies.

# Acknowledgments

This work was supported by the National Key R&D Program of China (No. 2022ZD0115002). The authors thank Dr. Yi Isaac Yang and Yifan Li for useful discussion. J.Z. also thanks the Tiger supercomputer cluster in Princeton University.

# Additional Information

**Conflict of Interest**

The authors declare no conflict of interest.

**Code Availability**

Codes of CCG and associated model checkpoints will be released upon publication.

# References

1. Frenkel, D.; Smit, B., *Understanding molecular simulation: from algorithms to applications*. Elsevier: 2001; Vol. 1.

2. Monticelli, L.; Kandasamy, S. K.; Periole, X.; Larson, R. G.; Tieleman, D. P.; Marrink, S.-J., The MARTINI coarse-grained force field: extension to proteins. *Journal of chemical theory and computation* **2008,** *4* (5), 819-834.

3. Saunders, M. G.; Voth, G. A., Coarse-graining methods for computational biology. *Annual review of biophysics* **2013,** *42*, 73-93.

4. Fiorin, G.; Klein, M. L.; Hénin, J., Using collective variables to drive molecular dynamics simulations. *Molecular Physics* **2013,** *111* (22-23), 3345-3362.

5. Noid, W.; Chu, J.-W.; Ayton, G. S.; Krishna, V.; Izvekov, S.; Voth, G. A.; Das, A.; Andersen, H. C., The multiscale coarse-graining method. I. A rigorous bridge between atomistic and coarse-grained models. *The Journal of chemical physics* **2008,** *128* (24), 244114.

6. Torrie, G. M.; Valleau, J. P., Nonphysical sampling distributions in Monte Carlo free-energy estimation: Umbrella sampling. *Journal of Computational Physics* **1977,** *23* (2), 187-199.

7. Abrams, C.; Bussi, G., Enhanced sampling in molecular dynamics using metadynamics, replica-exchange, and temperature-acceleration. *Entropy* **2014,** *16* (1), 163-199.

8. Okamoto, Y., Generalized-ensemble algorithms: enhanced sampling techniques for Monte Carlo and molecular dynamics simulations. *Journal of Molecular Graphics and Modelling* **2004,** *22* (5), 425-439.

9. Yang, Y. I.; Shao, Q.; Zhang, J.; Yang, L.; Gao, Y. Q., Enhanced sampling in molecular dynamics. *The Journal of Chemical Physics* **2019,** *151* (7), 070902.

10. Reith, D.; Pütz, M.; Müller‐Plathe, F., Deriving effective mesoscale potentials from atomistic simulations. *Journal of computational chemistry* **2003,** *24* (13), 1624-1636.

11. Izvekov, S.; Voth, G. A., A multiscale coarse-graining method for biomolecular systems. *The Journal of Physical Chemistry B* **2005,** *109* (7), 2469-2473.

12. Shell, M. S., The relative entropy is fundamental to multiscale and inverse thermodynamic problems. *The Journal of chemical physics* **2008,** *129* (14), 144108.

13. Schneider, E.; Dai, L.; Topper, R. Q.; Drechsel-Grau, C.; Tuckerman, M. E., Stochastic neural network approach for learning high-dimensional free energy surfaces. *Physical review letters* **2017,** *119* (15), 150601.

14. Wang, J.; Olsson, S.; Wehmeyer, C.; Pérez, A.; Charron, N. E.; De Fabritiis, G.; Noé, F.; Clementi, C., Machine learning of coarse-grained molecular dynamics force fields. *ACS central science* **2019**.

15. Zhang, J.; Lei, Y.-K.; Yang, Y. I.; Gao, Y. Q., Deep learning for variational multiscale molecular modeling. *The Journal of Chemical Physics* **2020,** *153* (17), 174115.
16. Köhler, J.; Chen, Y.; Krämer, A.; Clementi, C.; Noé, F., Force-matching Coarse-Graining without Forces. *arXiv preprint arXiv:2203.11167* **2022**.
17. Yang, Y. I.; Gao, Y. Q., Computer simulation studies of Aβ37–42 aggregation thermodynamics and kinetics in water and salt solution. *The Journal of Physical Chemistry B* **2015,** *119* (3), 662-670.
18. Johnston, K.; Harmandaris, V., Hierarchical simulations of hybrid polymer–solid materials. *Soft Matter* **2013,** *9* (29), 6696-6710.
19. Weinan, E.; Engquist, B., The heterognous multiscale methods. *Communications in Mathematical Sciences* **2003,** *1* (1), 87-132.
20. Heath, A. P.; Kavraki, L. E.; Clementi, C., From coarse‐grain to all‐atom: toward multiscale analysis of protein landscapes. *Proteins: Structure, Function, and Bioinformatics* **2007,** *68* (3), 646-661.
21. Wassenaar, T. A.; Pluhackova, K.; Böckmann, R. A.; Marrink, S. J.; Tieleman, D. P., Going backward: a flexible geometric approach to reverse transformation from coarse grained to atomistic models. *Journal of chemical theory and computation* **2014,** *10* (2), 676-690.
22. Lombardi, L. E.; Martí, M. A.; Capece, L., CG2AA: backmapping protein coarse-grained structures. *Bioinformatics* **2016,** *32* (8), 1235-1237.
23. Chen, L.-J.; Qian, H.-J.; Lu, Z.-Y.; Li, Z.-S.; Sun, C.-C., An automatic coarse-graining and fine-graining simulation method: Application on polyethylene. *The Journal of Physical Chemistry B* **2006,** *110* (47), 24093-24100.
24. Schlitter, J.; Engels, M.; Krüger, P., Targeted molecular dynamics: a new approach for searching pathways of conformational transitions. *Journal of molecular graphics* **1994,** *12* (2), 84-89.
25. Yang, W.; Zhang, X.; Tian, Y.; Wang, W.; Xue, J.-H.; Liao, Q., Deep learning for single image super-resolution: A brief review. *IEEE Transactions on Multimedia* **2019,** *21* (12), 3106-3121.
26. Li, W.; Burkhart, C.; Polińska, P.; Harmandaris, V.; Doxastakis, M., Backmapping coarse-grained macromolecules: An efficient and versatile machine learning approach. *The Journal of Chemical Physics* **2020,** *153* (4), 041101.
27. Noé, F.; Olsson, S.; Köhler, J.; Wu, H., Boltzmann generators: Sampling equilibrium states of many-body systems with deep learning. *science* **2019,** *365* (6457).
28. Goodfellow, I.; Pouget-Abadie, J.; Mirza, M.; Xu, B.; Warde-Farley, D.; Ozair, S.; Courville, A.; Bengio, Y. In *Generative adversarial nets*, Advances in neural information processing systems, 2014; pp 2672-2680.
29. Kingma, D. P.; Welling, M., Auto-encoding variational bayes. *arXiv preprint arXiv:1312.6114* **2013**.
30. Blei, D. M.; Kucukelbir, A.; McAuliffe, J. D., Variational inference: A review for statisticians. *Journal of the American Statistical Association* **2017,** *112* (518), 859-877.
31. Sohl-Dickstein, J.; Weiss, E.; Maheswaranathan, N.; Ganguli, S. In *Deep unsupervised learning using nonequilibrium thermodynamics*, International Conference on Machine Learning, PMLR: 2015; pp 2256-2265.
32. Ho, J.; Jain, A.; Abbeel, P., Denoising diffusion probabilistic models. *Advances in Neural Information Processing Systems* **2020,** *33*, 6840-6851.
33. Hills Jr, R. D.; Lu, L.; Voth, G. A., Multiscale coarse-graining of the protein energy landscape. *PLoS computational biology* **2010,** *6* (6), e1000827.
34. Abdolmaleki, A.; Springenberg, J. T.; Degrave, J.; Bohez, S.; Tassa, Y.; Belov, D.; Heess, N.; Riedmiller, M., Relative entropy regularized policy iteration. *arXiv preprint arXiv:1812.02256* **2018**.
35. Lemaréchal, C., Lagrangian relaxation. In *Computational combinatorial optimization*, Springer: 2001; pp 112-156.
36. Zhu, J.-Y.; Park, T.; Isola, P.; Efros, A. A. In *Unpaired image-to-image translation using cycle-consistent adversarial networks*, Proceedings of the IEEE international conference on computer vision, 2017; pp 2223-2232.
37. Chen, X.; Duan, Y.; Houthooft, R.; Schulman, J.; Sutskever, I.; Abbeel, P., Infogan: Interpretable representation learning by information maximizing generative adversarial nets. *Advances in neural information processing systems* **2016,** *29*.

38. Sugita, Y.; Okamoto, Y., Replica-exchange molecular dynamics method for protein folding. *Chemical physics letters* **1999,** *314* (1-2), 141-151.
39. Gao, Y. Q., An integrate-over-temperature approach for enhanced sampling. *The Journal of chemical physics* **2008,** *128* (6), 064105.
40. Laio, A.; Parrinello, M., Escaping free-energy minima. *Proceedings of the National Academy of Sciences* **2002,** *99* (20), 12562-12566.
41. Zhang, L.; Wang, H.; E, W., Reinforced dynamics for enhanced sampling in large atomic and molecular systems. *The Journal of chemical physics* **2018,** *148* (12), 124113.
42. Weinan, E.; Ren, W.; Vanden-Eijnden, E., Finite temperature string method for the study of rare events. *J. Phys. Chem. B* **2005,** *109* (14), 6688-6693.
43. Zhang, J.; Lei, Y.-K.; Zhang, Z.; Han, X.; Li, M.; Yang, L.; Yang, Y. I.; Gao, Y. Q., Deep reinforcement learning of transition states. *Physical Chemistry Chemical Physics* **2021,** *23* (11), 6888-6895.
44. Tiwary, P.; Berne, B., Predicting reaction coordinates in energy landscapes with diffusion anisotropy. *The Journal of chemical physics* **2017,** *147* (15), 152701.
45. Grathwohl, W.; Chen, R. T. Q.; Bettencourt, J.; Sutskever, I.; Duvenaud, D., FFJORD: Free-Form Continuous Dynamics for Scalable Reversible Generative Models. In *International Conference on Learning Representations*, 2019.
46. Zwanzig, R., *Nonequilibrium statistical mechanics*. Oxford University Press: 2001.
47. Honda, S.; Yamasaki, K.; Sawada, Y.; Morii, H., 10 residue folded peptide designed by segment statistics. *Structure* **2004,** *12* (8), 1507-1518.
48. Lindorff-Larsen, K.; Piana, S.; Dror, R. O.; Shaw, D. E., How fast-folding proteins fold. *Science* **2011,** *334* (6055), 517-520.
49. Behler, J., Perspective: Machine learning potentials for atomistic simulations. *journal of chemical physics* **2016,** *145* (17).
50. Bartók, A. P.; De, S.; Poelking, C.; Bernstein, N.; Kermode, J. R.; Csányi, G.; Ceriotti, M., Machine learning unifies the modeling of materials and molecules. *Science advances* **2017,** *3* (12), e1701816.
51. Zhang, L.; Han, J.; Wang, H.; Car, R.; Weinan, E., Deep potential molecular dynamics: a scalable model with the accuracy of quantum mechanics. *Physical review letters* **2018,** *120* (14), 143001.
52. Faradjian, A. K.; Elber, R., Computing time scales from reaction coordinates by milestoning. *The Journal of chemical physics* **2004,** *120* (23), 10880-10889.
53. Song, Y.; Sohl-Dickstein, J.; Kingma, D. P.; Kumar, A.; Ermon, S.; Poole, B., Score-based generative modeling through stochastic differential equations. *arXiv preprint arXiv:2011.13456* **2020**.
54. Lipman, Y.; Chen, R. T.; Ben-Hamu, H.; Nickel, M.; Le, M., Flow matching for generative modeling. *arXiv preprint arXiv:2210.02747* **2022**.
55. Wu, H.; Mardt, A.; Pasquali, L.; Noe, F. In *Deep generative markov state models*, Advances in Neural Information Processing Systems, 2018; pp 3975-3984.
56. Sidky, H.; Chen, W.; Ferguson, A. L., Molecular latent space simulators. *Chemical Science* **2020,** *11* (35), 9459-9467.